\documentstyle[12pt]{article}
\input{epsf}
\newcommand{\be}{\begin{eqnarray}}
\newcommand{\ee}{\end{eqnarray}}

\newcommand{\ha}{\frac{1}{2}}

\newcommand{\gd}{\delta}

\newcommand{\gth}{\theta}

\newcommand{\gs}{\sigma}

\newcommand{\gff}{\varphi}

\newcommand{\gD}{\Delta}

\newcommand{\gL}{\Lambda}

\newcommand{\gS}{\Sigma}

\newcommand{\gW}{\Omega}

\newcommand{\ti}[1]{\tilde{#1}}

\newcommand{\ov}{\overline}
\newcommand{\ra}{\rightarrow}
\newcommand{\lan}{\langle}
\newcommand{\ran}{\rangle}
\newcommand{\noi}{\noindent}

\begin{document}
{\hbox to\hsize{}
\begin{center}
{\Large \bf {Power filtration of CMB observational data} }\\
\bigskip
{\bf D.I. Novikov
\footnote{Also: Astro - Space Center of Lebedev Physical Institute,
Profsoyuznaya 84/32, Moscow, Russia.\\
University Observatory, Juliane Maries Vej 30, 2100 Copenhagen, Denmark.\\
Ludwig Maximilians Universit{\"a}t,
Theresienstrasse 37, D-80333 M{\"u}nchen, Germany}\\
{\it Astrophysics Department of Physics, Keble Road, Oxford OX1 
3RH, U.K.}}\\
{\bf P. Naselsky
\footnote{Also: Theoretical Astrophysics Center
Juliane Maries Vej 30, DK-2100, Copenhagen, Denmark}\\
{\it  Rostov State University, Zorge 5, 344090 Rostov-Don, Russia}}\\
{\bf H.E. J{\o}rgensen\\
{\it Copenhagen University Observatory, Juliane Maries Vej 30,
DK-2100 Copenhagen, Denmark.}}\\
{\bf P.R. Christensen
\footnote{Also: Niels Bohr Institute, Blegdamsvej 17, DK-2100 
Copenhagen, Denmark}\\
{\it Theoretical Astrophysics Center, Juliane Maries Vej 30,
DK-2100 Copenhagen, Denmark}}\\
{\bf I.D. Novikov
\footnote{ Also: Astro - Space Center of Lebedev Physical Institute,
Profsoyuznaya 84/32, Moscow, Russia; University Observatory,
Juliane Maries Vej 30, DK-2100, Copenhagen;
NORDITA, Blegdamsvej 17, DK-2100, Copenhagen, Denmark.}\\
{\it Theoretical Astrophysics Center, Juliane Maries Vej 30, 2100
Copenhagen, {\O} Denmark}}\\
{\bf H.U. N\o rgaard-Nielsen\\
{\it  Danish Space Research Institute, Juliane Maries Vej 30,
DK-2100 Copenhagen, Denmark}} \\
\date{\today}
\end{center}
\begin{abstract}

We propose a power filter $G_p$ for linear reconstruction of the CMB 
signal from one-dimensional scans of observational maps. This $G_p$ 
filter preserves the power spectrum of the CMB signal in contrast to 
the Wiener filter which diminishes the power spectrum of the 
reconstructed CMB signal. We demonstrate how peak statistics and a 
cluster analysis can be used to estimate the probability of the presence 
of a CMB signal in observational records. The efficiency of the $G_p$ 
filter is demonstrated on a toy model of an observational record 
consisting of a CMB signal and noise in the 
form of foreground point sources.
\end{abstract}

\section{Introduction}

The exploration of anisotropy and polarization of the Cosmic Microwave 
Background (CMB) is fundamental in developing our knowledge about 
the Universe.
During the last ten years several CMB measurement projects are being
carried out $^1$. After 
the successful COBE mission the attention has been focused on 
obtaining higher angular resolution of observational data in the range
of the so-called Doppler peak and subsequent peaks. 
The level of the CMB anisotropy at multipoles
$l>30$ up to $l\sim 10^3 -  2\cdot10^3$ 
is a gold mine of cosmological information about the early Universe and 
its most important  parameters.
However, statistical analysis of modern observational data is 
extremely complicated and time consuming. This complexity leads to 
development of different approximate methods of the data reduction such 
as Wiener-filtration $^{2,3}$, radical compression method $^4$, likelihood 
method,  band-power method and others $^{5,6,7}$.
The existence of a large number of  processing methods of 
observational data for different types of experiments is easily understood.
Firstly, there are many difficulties and peculiarities of the
measurements connected with the scan strategy, specific beam chopping, 
map making, etc.
Secondly, the reconstruction of the parameters of the CMB  signal
is a sort of ``inverse problem''.                
These difficulties are not specific for CMB measurements only. They are 
well known also in optical astronomy and in image reconstruction methods
from satellite or airplane photography $^8$.

Which one of the data reduction methods mentioned above is the most 
preferable for 
any type of experiments? The answer of this question depends on the our 
``intuitive'' hopes. We can use the following criterion which was 
suggested by Tegmark $^{2}$: ``one method is better than another if it 
retains more of the cosmological information which operationally means 
that it will lead to smaller error bars on the parameter estimates''.
However, before CMB image reconstruction from the observational 
scans is done we have no information about the structure of the signal 
(systematic 
errors, Gaussianity, different types of foreground sources and noise, etc). 
So, roughly speaking, the Tegmark criterion plays an important role 
a posteriori, when the reconstruction of the CMB by different methods is 
performed. 
We have to remember that for different strategies of observations and for
different concrete experiments different methods of data processing can
be considered as the most preferable. In addition 
%for different aims of using of obtained data 
different methods are preferable for different goals for using obtained 
data. 
In general, we may conclude that different methods complement each other.

The purpose of this paper is to propose and investigate a new method with
the following aim: how can we estimate the probability of the presence of 
the CMB signal in an observational map and determine its characteristics
if we have a hypothesis about its statistical properties
(for example that it is Gaussian), and have a hypothesis about its power
spectrum.

We suggest a new type of linear filtration of one-dimensional scans of CMB 
maps that is sensitive to non-Gaussian noise of any origin. This filtration 
does not change the primordial CMB power spectrum in contrast to the well 
known Wiener filter method.
We use geometrical and topological characteristics such as the Minkowski 
functionals $^{9-11}$ and statistics of peak distribution above and below 
some definite threshold $^{12-15}$.
We show, that these characteristics are very sensitive criteria of 
Gaussianity of the restored signal.
The filtration of the two-dimensional maps of the CMB with the filter
which preserves the statistical properties of the underlying signal 
was described in $^7$.

The full advantage of the new method can be demonstrated even on very
small data sets. We apply our method to  a ``toy'' model of a mixture
of CMB signal and foreground point sources to model
balloon- and ground-based measurements at the frequency band 
$\nu\simeq 10 - 200$GHz.
The application of the method to real observational data will be 
considered in a separate paper.

The rest of this paper is organized as follows. In Section 2 we describe 
the model of the real experiments and describe a new so called 
``power filter'' for reconstructing the CMB signal. 
Section 3 is devoted to peak statistics and cluster analysis 
of the CMB signal in one-dimensional records. In Section 4 we present the 
new reconstruction technique of the CMB spectrum from records with 
point sources and demonstrate the strength of our method. 
Section 5 contains our conclusions.

\section{The model of real experiments and reconstruction of the CMB signal}

The statistically isotropic distribution of the CMB temperature fluctuations 
on the sky is usually described by the spherical harmonics expansion 
$Y_{lm}(\vec{q})$: 
\be
T(q)=\sum_{l=1}^{\infty}\sum_{m=-l}^{l}a_{lm}B_{lm} Y_{lm}
(\vec{q}),
\label{1}
\ee
where $a_{lm}$ are zero-mean Gaussian deviates, 
$\lan a_{lm}a^*_{l'm'}\ran=C_l \gd_{ll'}\gd_{mm'}$,
with $C_l$ given by the CMB power spectrum, $B_{lm}$ the antenna beam, 
and $\vec{q}(\gth,\gff)$ the unit vector defining the position of the 
observational  direction on the sky. Eq.(\ref{1}) can be written for all 
frequency channels $\nu_1...\nu_{\ti{m}}$, where $\ti{m}$ is the number 
of bands. After the COBE mission, different CMB experiments such as Saskatoon,
QMAP, CAT and others have observed rather small parts of the sky with  higher 
angular resolution. They estimated statistical characteristics of the CMB
signal from the maps using some specific pixelization strategy. Two future 
space missions, MAP and PLANCK, will cover a significantly greater part of the 
sky and will provide data with higher precision. 

The traditional view is that most information of a Gaussian random field 
is contained in the multipole power spectrum $C_l$. As a result a great 
deal of efforts has gone into obtaining the best estimate of the power 
spectrum $C_l$ from the experimental data. The power spectrum $C_l$ 
describes properties of a two-dimensional map.
However, there exist many balloon-borne experiments and ground-based 
projects which performed observations of the CMB anisotropy 
along one-dimensional sky patches. Besides that the future MAP 
and PLANCK missions will collect data from a large number of intersecting 
circles which will then be reconstructed into a two-dimensional sky map.
Briefly speaking, because of the simplicity of data analysis in the
one-dimensional case, the collection of CMB observational data in this 
format is an interesting option for modern  anisotropy 
experiments.

Usually (see ref. $^2$ for a recent review) one has a time-ordered set of 
data and can, knowing the adopted scan strategy, 
obtain a map with pixel values denoting the sum of a
pure CMB signal ${\mathbf{x}}$ and different types ${\mathbf{n}}$ of noise, 
including foregrounds, atmospheric emission, radiometer noise and so on. 

In the following we consider final maps consisting of a pure CMB signal
${\mathbf{ x}}$ and noise ${\mathbf{n}}=\sum_i {\mathbf{n}}_i$ and use therefore
for a one-dimensional single record 
\be
{\mathbf{Y(t)}}={\mathbf{x}}(t)+{\mathbf{n}}(t)
\label{2}
\ee
where $t$ is an independent variable along the record.

For this simple model of the measured signal Eq.(\ref{2}) one can 
approximately reconstruct the  pure CMB signal using a linear 
reconstruction technique (see $^8$ for a general description of the 
method). Note that in Eq.(\ref{2}) we do not assume  the random fields 
${\mathbf{x}}$ and ${\mathbf{n}}$ to be Gaussian.  

The general task of linear methods of reconstruction of a signal is 
to obtain an approximate estimation $\ov{{\mathbf{x}}}'(t)$ of a signal 
${\mathbf{x}}(t)$ from the measured signal ${\mathbf{Y(t)}}$ using
linear operations. Following $^8$ the estimation $\ov{{\mathbf{x}}}'(t)$ of 
the CMB signal ${\mathbf{x}}(t)$ from Eq.(\ref{2}) can be written as
\be
\ov{{\mathbf{x}}}'(t)=\int G(t-t'){\mathbf{Y}}(t')dt'
\label{3}
\ee
where $G(t-t')$ is the filter of the reconstruction.

Before describing the new method which we propose for the reconstruction of 
the CMB signal, let us briefly discuss the standard linear filtering 
methods which were reviewed by Tegmark $^2$. Firstly, following  $^2$, we 
point out that linear methods of  filtration of maps
do not destroy information about phases which is contained in the initial 
maps (or one-dimensional records). Secondly, the non-linear extraction of the 
signal, such as band-power method $^{4,5}$, needs some special assumptions 
about the statistical properties of CMB signal and noise (for example 
Gaussian statistics of noise).
Besides that, all non-linear methods destroy information contained 
in the maps $^2$.

Let us return to  Eq.(\ref{2}) and Eq.(\ref{3}) and consider one example 
of CMB signal reconstruction which uses a simple Wiener filtering 
method $^2$. Following $^2$, the estimated signal from Eq.(\ref{2}) is
\be
\ov{{\mathbf{x}}}''(t)=\int W(t-t'){\mathbf{Y}}(t)dt,
\label{4}
\ee
where $W(t-t')$ is the so called Wiener filter which minimizes the
reconstruction error 
$\gD^2\equiv (\ov{{\mathbf{x}}}''-{\mathbf{x}})^2$. Tegmark and Efstathiou 
$^3$ and Tegmark $^2$ have shown that the W filter has a particularly
simple form in Fourier space under assumptions in
Eq.(\ref{2}). For this case $W$ has the following form:
\be
W(k)=\frac{P_{CMB}(k)}{P_{tot}(k)}=\left(1+
\frac{P_{noise}(k)}{P_{CMB}(k)}\right)^{-1},
\label{5}
\ee
where $P_{tot}(k)=P_{CMB}(k)+P_{noise}(k)$ is the power spectrum 
of the measured signal ${\mathbf{Y}}(t)$, $P_{CMB}(k)$ the CMB power 
spectrum
folded with the antenna beam and $P_{noise}(k)$ the noise power spectrum.

Let us return to Eq.(\ref{4}). 
It is straightforward to prove that the power spectrum of the 
reconstructed signal $P_{\ov{x}''}(k)$ is equal to
\be
P_{\ov{x}''}(k)=W^2(k)P_{tot}(k)=P_{CMB}(k)W(k).
\label{6}
\ee
As one can see from Eq.(\ref{5}), $W<1$ which, together with 
Eq.(\ref{6}), shows that the Wiener filter decreases the CMB power spectrum 
of the reconstructed signal.

For a linear reconstruction of the CMB signal in one-dimensional scans we 
propose to use the so called ``power filter'' $G_p$ $^8$. As we told in 
Section 1 the corresponding consideration for two-dimensional CMB maps was 
done in $^7$. This filter minimizes the difference between the power spectrum 
$P_{\ov{x}'}(k)$ and the CMB power spectrum.
It is clear that this condition corresponds to 
\be
P_{\ov{x}'}(k)=P_{CMB}.
\label{7}
\ee
We will demonstrate that for our case $G_p=W^{1/2}$. For the proof
let us rewrite  Eq.(\ref{3}) for our filter $G_p$:
\be
\ov{{\mathbf{x}}}'(t)=\int G_p(t-t'){\mathbf{Y}}(t')dt'.
\label{8}
\ee
Now note that from Eq.(\ref{8}) one can conclude that 
\be
P_{x'}(k)=G_p^2(k)P_{tot}(k).
\label{9}
\ee
Using Eq.(\ref{5}), Eq.(\ref{7}) and Eq.(\ref{9}) we conclude that
\be
 G^2_p(k)=\frac{P_{CMB}(k)}{P_{tot}(k)}=W(k)
\label{10}
\ee
which proves our statement: $G_p=W^{1/2}$.

\section{Peak statistics and cluster analysis of the reconstructed 
CMB signal}

Using power filtration by a $G_p$ filter we transform  the initial 
observational map to a reconstructed record
$\{\ov{\mathbf{x}'}(t)\}$. Like  any other linear filter the $G_p$ filter
transforms the power spectrum of the initial signal preserving
phases of the signal. This property of the linear filtration is 
extremely important and we will consider it in more details.
Let us consider a one-dimensional record $Y(t)$ and its Fourier expansion:
\be 
Y(t)=\sum^{\infty}_{n=1} 
\left[a_n\cos\left(\frac{2\pi nt}{\ti{T}}\right)+
b_n\sin\left(\frac{2\pi nt}{\ti{T}}\right)\right],
\label{11}
\ee
where $\ti{T}$ is the total length of a record. After a power filtration we 
get
\be
Y'(t)=\sum^{\infty}_{n=1} 
\left[a_nG_n\cos\left(\frac{2\pi nt}{\ti{T}}\right)+
b_nG_n\sin\left(\frac{2\pi nt}{\ti{T}}\right)\right].
\label{12}
\ee
The initial power spectrum of the record (Eq.(11)) is given by
\be
P_n^{(in)}=\ha (a^2_n+b^2_n)
\label{13}
\ee
and of the reconstructed power spectrum (Eq.(12)) by 
\be
P_n^{(rec)}=\ha G^2_n(a^2_n+b^2_n).
\label{14}
\ee
However our $G_p$ filter does not destroy the spectrum of phases since
\be
\Phi_n^{(in)}=\arctan\left(\frac{b_n}{a_n}\right)=\Phi_n^{(rec)}.
\label{15}
\ee
So, roughly speaking, we transform 
the power spectrum of the initial signal by $G_p$ filter but preserve the 
spectrum of phases. Thus  Eq.(\ref{15}) demonstrates that such a 
filtration preserves a Gaussian or non-Gaussian property of the modes which
are not suppressed by the filter.
This property is valid for all linear filters. The specific 
property of the $G_p$ filtration is that in addition to preserve the 
phases it also preserves the spectrum of the CMB signal. $G_p$ is a 
unique filter with this property.
Cluster analysis and peak statistics $^{11-14}$ are very sensitive to both 
phases and the power spectrum. Thus, if we believe that after the 
$G_p$ filtration the reconstructed signal really represents the 
approximate distribution of $\gD T$ of CMB signal, it should have the 
same phase characteristics and the same peak statistics and properties 
of clusterisation which are predicted for the initial CMB signal (under 
an accepted hypothesis about the cosmological model). Confrontation of peak
statistics and clusterisation of the reconstructed signal with the 
theoretical predictions is a good test of the ``quality'' of the 
reconstruction.

We will investigate how this method works in the case of one-dimensional 
records.  In the previous section we have shown that the 
reconstructed record ${\mathbf{  x'}}(t)$ has the power spectrum 
$P_{CMB}$. We would like to point out that all non-Gaussian features 
of the initial signal are transformed to the reconstructed 
record ${\mathbf{  x'}}(t)$ because of the linear character of filtration.
We also note that in the case of application of the Wiener filter, peak
statistics of the reconstructed signal is different from the hypothetical 
CMB signal because of the change of the power spectrum.

Assuming the primordial cosmological signal to be
Gaussian, we perform the test of the filtered signal with regard to 
Gaussianity. In fact, we calculate the probability that the filtered 
signal has a Gaussian nature.

The $G_p$ filtering obviously makes the power spectrum of the observed
signal equal to the expected CMB spectrum. The rest of the 
information to be investigated is about the phases only. In order to
perform this test, we use a geometrical approach.

We normalize the ${\mathbf{  x'}}(t)$ record to its variance: 
$\nu(t)\equiv \gD T/\sqrt{\lan \gD T^2\ran}$ and consider the 
following characteristics of the $\nu(t)$ statistics:  

{\bf A)} Normalized distribution functions
of  maxima and minima above the threshold  $\nu_t$:
\be
f_{max}(\nu_t)=\frac{N_{max}(\nu_t)}{\ov{N_{max}}} \hspace{1cm}
f_{min}(\nu_t)=\frac{N_{min}(\nu_t)}{\ov{N_{min}}},
\label{16}
\ee
where $N_{max}(\nu_t)$ and $N_{min}(\nu_t)$ are the number of 
maxima and minima, respectively, and  
$\ov{N_{max}}=N_{max}(\nu_t\ra-\infty)$, 
$\ov{N_{min}}=N_{min}(\nu_t\ra-\infty)$ are the total number of 
maxima and minima.

{\bf B)} The global Minkowski functionals $M_1$ and  $M_2$  for 
a one-dimensional record, which 
characterize the cumulative distribution function $(M_1)$ and the Euler 
characteristic (genus) of the random signal $(M_2)$. 
The last quantity was first introduced in cosmology by Doroshkevich $^{16}$.
For the N dimensional case there are N+1 Minkowski functionals.

{\bf C)} Finally we consider the mean length of the clusters above and
below a constant threshold $\nu_t$.
A one-dimensional cluster of maxima is the 
continuous part of the curve with $\nu(t)>\nu_t$ inside the 
interval $t\in(t_1,t_2)$. Here $t_1$ and $t_2$ are roots of the equation 
$\nu(t_{1,2})=\nu_t$ and $t_2>t_1$. The length $k(\nu_t)$ of the 
cluster is defined as the number of 
peaks with height $\nu_{peak}>\nu_t$ in the cluster.
If the value of $\nu_t$ is high ($\nu_t\gg 1$) then only high 
maxima (separated from each other by regions with $\nu(t)<\nu_t$) 
are present above the threshold and the typical
length of a cluster is $\lan k(\nu_t)\ran=1$.
The reduction of the threshold level $\nu_t$ down to $\nu_t\ra0$ 
or $\nu_t<0$ leads to the appearance of big clusters when maxima of 
smaller clusters begin to connect together and generate a new cluster. 
This process is characterized by the number of clusters $N_k(\nu_t)$ 
of the length $k$ and the total number of clusters $N(\nu_t)$ which 
are present in the anisotropy records for an appropriate threshold 
$\nu_t$:
\be
N(\nu_t)=\gS^{\infty}_{k=1}N_k(\nu_t).
\label{17}
\ee
Thus for different statistics of initial records the mean length of 
a cluster at threshold level $\nu_t$ is
\be
\lan k(\nu_t)\ran=
\frac{\gS^{\infty}_{k=1}kN_k(\nu_t)}
{\gS^{\infty}_{k=1}N_k(\nu_t)}.
\label{18}
\ee

It is necessary to note, that only the first characteristic
considered under {\bf A} and the first Minkowski functional  $M_1$ are 
independent. The rest of them can be represented in terms of the 
distribution of peaks ({\bf A}). Thus, we use them only for visual 
clarification and representation of our results.

The defining Eq.(\ref{16}) - Eq.(\ref{18}) do not depend 
on the nature of the initial signal or its composition.

Below we use the standard definition of the spectral parameters:

\be
\begin{array}{l}
\sigma_n^2=\int k^{2n}P_{CMB}(k)dk, \hspace{3cm} n=0,1,2 \\
\gamma=\frac{\sigma_1^2}{\sigma_0\sigma_2}
\end{array}
\label{19}
\ee   

where $P_{CMB}(k)$ is the one-dimensional CMB power spectrum. 

The characteristics {\bf A, B} and {\bf C} have a rather simple form in case of 
a random Gaussian field $^{10-12,14}$:
 
The number of maxima and minima above a threshold $\nu_t$ are:

\be
\begin{array}{l}
N_{max}(\nu_t)=\frac{1}{4\pi}\frac{\sigma_2}{\sigma_1}
\left( 1-\Phi\left(\frac{\nu_t}{\sqrt{2(1-\gamma^2)}}\right)+
\gamma e^{-\frac{\nu_t^2}{2}}\left( 1+\Phi\left(\frac{\gamma\nu_t}
{\sqrt{2(1-\gamma^2)}}\right)\right)\right) \\

N_{min}(\nu_t)=\frac{1}{4\pi}\frac{\sigma_2}{\sigma_1}
\left( 1-\Phi\left(\frac{\nu_t}{\sqrt{2(1-\gamma^2)}}\right)-
\gamma e^{-\frac{\nu_t^2}{2}}\left( 1-\Phi\left(\frac{\gamma\nu_t}
{\sqrt{2(1-\gamma^2)}}\right)\right)\right) \\

\ov{N_{max}}=\ov{N_{min}}=\frac{1}{2\pi}\frac{\sigma_2}{\sigma_1} \\

\end{array}
\label{20}
\ee   
The Global Minkowski functionals for a one-dimensional Gaussian field are given by:
\be
\begin{array}{l}
M_1(\nu_t)=\ha\left(1-\Phi(\frac{\nu_t)}{\sqrt{2}})\right)\\
M_2(\nu_t))=f_{max}(\nu_t)-f_{min}(\nu_t)=
\gamma e^{-\frac{\nu_t^2}{2}},\\
\end{array}
\label{21}
\ee
where $\Phi(x)=\frac{2}{\sqrt{\pi}}\int^x_0 e^{-t^2}dt$.

As it has been already mentioned above, the last characteristic ($M_2$)
can be written in terms of the peak distributions ({\bf A}) $^{14}$:
\be
\lan k(\nu_t)\ran=\frac{N_{max}(\nu_t)}{N_{max}(\nu_t)-N_{min}(\nu_t)}
\label{22}
\ee
It is not difficult to demonstrate that for a Gaussian distribution all 
these characteristics are functions of only two variables:
threshold $\nu_t$ and spectral parameter $\gamma$. This fact makes
our analysis transparent. 

To calculate the probability that the restored (filtered) signal is
Gaussian, we use only the characteristic as mentioned under {\bf A}.
The distributions of maxima  and minima, 
$f_{max}(\nu_t,\gamma)$ and $f_{min}(\nu_t,\gamma)$, above a threshold $\nu_t$
are monotonic functions  of $\nu_t$ with
\be
\begin{array}{l}
f_{max}(-\infty,\gamma)=f_{min}(-\infty,\gamma)=1 \\
f_{max}(+\infty,\gamma)=f_{min}(+\infty,\gamma)=0 .\\
\end{array}
\label{23}
\ee
For a limited number of peaks these distributions look like a 
step functions with number of steps equal to the total
number of maxima and minima. If the observed signal has a cosmological nature,
then this function should be consistent with the smooth analytical
curve (see Eq. (16) and (20)) for $\gamma$ of the CMB spectrum. In  
case of real observations in a limited record we deal
with bad statistics (small number of peaks). For this kind of
statistics we have to perform an appropriate test, e.g. a 
Kolmogorov-Smirnov test $^{17}$. 
Before applying such a test and to demonstrate the advantage of our method, we
should do the following:\\
1. As representations of the spectrum of the filtered signal (being identical 
to the spectrum of the hypothetical CMB signal) we simulate a large 
number $n_r$ of different realizations of a random Gaussian field in the 
considered record using a FFT algorithm.\\
2. For each realization r=1,..,$n_r$ we calculate $\gamma_r$
using the definition (19) as well as $f^r_{max}$ and $f^r_{min}$ using Eq.(16). 
Furthermore we calculate the quantities:
\be
\begin{array}{l}
I_{max,real}(r,\gamma_r)=
\int_{-5}^{+5}f_{max}^r(\nu_t,\gamma_r)d\nu_t\\
I_{max,anal}(r,\gamma_r)=
\int_{-5}^{+5}f_{max}(\nu_t,\gamma_r)d\nu_t\\
I_{min,real}(r,\gamma_r)=
\int_{-5}^{+5}f_{min}^r(\nu_t,\gamma_r)d\nu_t\\
I_{min,anal}(r,\gamma_r)=
\int_{-5}^{+5}f_{min}(\nu_t,\gamma_r)d\nu_t,\\
\end{array}
\label{24}
\ee
and
\be
\begin{array}{l}
\gD_{max}=I_{max,real}-I_{max,anal}\\
\gD_{min}=I_{min,real}-I_{min,anal},\\
\end{array}
\label{25}
\ee
where $f_{max}(\nu_t,\gamma_r)$, $f_{min}(\nu_t,\gamma_r)$ have
the analytical form (20) and $f_{max}^r(\nu_t,\gamma_r)$, 
$f_{min}^r(\nu_t,\gamma_r)$ 
are distributions of maxima and minima in a given realization.
In (24) we performed integrations with the limits (-5,+5) where 
$f_{max}$ and $f_{min}$ in practice have reached their asymptotic 
values given by Eq.(23).
Obviously, $\gamma$ will differ slightly from one realization to another 
because of the cosmic variance. This difference
becomes smaller for longer records. Now we derive the formal probability 
that the distribution of peaks corresponds to a Gaussian distribution.
There are different methods to do this.  One of them is the following:\\  
3. We  count the fraction of realizations, where $|\gD_{max}(r)|$ and
$|\gD_{min}(r)|$ are greater than some positive value $z$
and calculate $P(z)$
\be
\begin{array}{l}
P_{max}(z)=\frac{N(|\gD_{max}(r)|>z)}{n_r}\\ 
P_{min}(z)=\frac{N(|\gD_{min}(r)|>z)}{n_r}\\
\end{array}
\label{26}
\ee
Here $P(z)$ represents a numerical estimate of how well the distribution of
maxima and minima corresponds to a Gaussian 
signal.\\
4. For a given filtered record we can now estimate to what extent the signal
has a Gaussian nature. We derive $\gD_{max}$ and $\gD_{min}$ to  
determine the probability $P(z)$.
It is clear that if $P(z)$ is larger than $\approx 3$\% and therefore
$z$ is within the $3 \gs$ level of the distribution, there is a 
good chance that we observe a Gaussian signal.
Alternatively one can calculate the differential functions $n(\gD )$ 
of the distributions of $\gD_{max}$ and $\gD_{min}$ and consider 
the actual value $n(\gD)$ for the 
filtered signal as a measure of closeness to a Gaussian signal. If this
$n(\gD)$ is greater than a few percent the filtered signal is Gaussian within
the "$3 \gs$ level" using the analogy with a standard Gaussian distribution. 

One could also use a Kolmogorov-Smirnov test, some of its generalizations 
or other methods to be discussed in a separate paper. 

In the next section we consider the application of this technique to a 
concrete toy model of an observed record.

\section{Application of the power filtration to a toy model of an 
observational record}

Now we demonstrate the application of the $G_p$ filtration 
and estimate to what extent we can recover the Gaussian nature of the 
signal. Furthermore we compare with the result from a Wiener filtration.

We consider a CMB spectrum of a standard Cold Dark Matter cosmological 
model with parameters: $h_{100}=0.75$, $\gW=1$, $\gW_b=0.0125$, 
$\gW_{\gL}=0$. For this spectrum we simulate $n_r$ = 1000 different realizations 
of the random Gaussian field (using a FFT algorithm) for a record with length 
$360^o$ and with FWHM of the antenna beam equal to $0.5^o$. 
Using the definitions in the previous section we calculate the values of 
$\gamma_r$, $I_{max}$, $I_{min}$ for each realization.
These values are slightly different from one record to another because of 
the cosmic variance. In Fig. 1 we show the 
dependence of $I_{max}$ and $I_{min}$ on $\gamma_r$ for the realizations.
In the same Fig. 1 we also show the corresponding Gaussian dependences.

Now we consider one specific realization (in analogy to a realization 
on the sky) which is marked by a star on Fig. 1 
($\gamma$= 0.3738, $I_{max}$= 2.751 and $I_{min}$= 2.271).
Next we compute for this realization $f(\nu_t)$ (see Fig. 2a), 
$M_1(\nu_t)$ (see Fig. 2b), rate of 
clusterisation ($N_{max}/(N_{max}-N_{min})$ and $N_{min}/(N_{min}-N_{max})$
as a function of $\nu_t$ (see Fig. 2c), and $M_2(\nu_t)$ (see Fig. 2d).
On Fig. 2(a-d) we also show the corresponding analytical dependence 
(see  Eq.(\ref{20}) and  Eq.(\ref{21})).
The graphs for the realization on Fig. 2(a-d) are only slightly different
from the corresponding analytical curves given by dashed lines.

Then we add to the CMB signal a signal from 100 identical 
foreground point-like sources with a random distribution over the 
record and with an amplitude such that the rms of the maps 
increases by a factor $\sqrt 2$ relative to the pure CMB signal. 
We consider the signal from the point-like sources as ``noise''.
The signal/noise ratio in this example is thus equal to 1.
Of course this is not a representation of a realistic noise but
only used to demonstrate the efficiency of the power filter $G_p$. 
Discussion of more realistic models will be given in a separate paper.  
The power spectrum of the CMB signal plus  noise is computed
for the same antenna beam with FWHM $0.5^o$. For this realization we 
determine $\gamma_{r}^{CMB+noise}$ and $I_{max}$ and $I_{min}$. 
The corresponding points are marked by crosses on Fig. 1 for the parameters
$\gamma$= 0.4833, $I_{max}$= 2.871 and $I_{min}$= 2.248.

It is seen that the crosses are far away from the clouds of realizations 
of the CMB and far from the Gaussian points (marked by dots with 
the same $\gamma_{r}^{CMB+noise}$ on Fig. 1). For this realization we 
calculated $f_{max}$, $f_{min}$, $M_1$, the rate of clusterisation, and 
$M_2$ and the corresponding analytical Gaussian functions for the given
$\gamma_{r}^{CMB+noise}$. The corresponding plots are presented on Fig. 3.
One can see that the distributions of these values on Fig. 3(a-d) differ 
significantly from the corresponding Gaussian functions. This means that 
the CMB+noise signal is far from being Gaussian. The same is seen from 
Fig.1. 
Now we perform filtration of this CMB+noise signal 
using the power filter $G_p$. As a result we expect to obtain a 
reconstructed signal 
with the power spectrum nearly identical to the CMB power spectrum of the 
considered realization. The corresponding points for $I_{max}$ 
and $I_{min}$ are indicated in Fig. 1 by circles with the parameters
$\gamma$= 0.3930, $I_{max}$= 2.782 and $I_{min}$= 2.272.

It is seen that the circles in fact are within the corresponding 
clouds of numerical realizations of the CMB signal and close to 
the corresponding Gaussian points with the same $\gamma$.

In Fig. 4 we have plotted $f_{max}$, $f_{min}$, $M_1$, the rate of 
clusterisation, and $M_2$ for the reconstructed signal as well as
the corresponding Gaussian analytical functions. One sees that 
the reconstructed realization is considerably closer to Gaussianity 
than the  CMB+noise signal, demonstrating how effective the power 
filter works. It is particularly important that the filtered signal 
has practically the same $\gamma$ as the input CMB signal. 
Furthermore we may get a  numerical estimate of how good the power 
filtration 
really is by using the method described in Section 3. The values of 
$n(\gD)$ for the filtered signal are $n(\gD_{max})$= 0.12 and 
$n(\gD_{min})$= 0.51.

For comparison we have performed a Wiener filtration of the 
same realization of CMB+noise signal and computed $\gamma^W$, 
$I_{max}^W$ and  $I_{min}^W$. The values
($\gamma$= 0.6232, $I_{max}$= 2.930 and $I_{min}$= 2.067)
are plotted as triangles on Fig. 1 
together with the Gaussian values for the same $\gamma^W$ as dots.

One can see that the Wiener filter displaces the corresponding points 
far away from the clouds of realizations of the CMB signal and that the 
power 
spectrum of the reconstructed signal is very far to be spectrum of the 
CMB signal.

\section{Conclusions}
In this paper we propose to use a power filter $^8$ for linear 
reconstruction of the CMB signal in one-dimensional scans. The specific 
property of the $G_p$ filter is that it preserves the power spectrum of 
the CMB signal. Under conditions 
described in Section 2 we get $G_p=W^{1/2}$ where $W$ is the
Wiener filter. Unlike the power filter $G_p$ the Wiener filter $W$ 
diminishes the power spectrum of the reconstructed signal compared to
the CMB power spectrum. In section 4 we have demonstrated that peak
statistics and cluster analysis can be used to estimate the efficiency
of the power filtration and to estimate  the
probability that a CMB signal is present in an observational record.
In Section 4 we also demonstrated how the power filter works using 
a toy model of an observational record consisting of the CMB signal and 
noise in the form of point like sources. We note that in the 
example of Section 4 we used a signal/noise ratio equal to unity
as described in section 4, 
but tests with different toy models have demonstrated that this 
method works well also in the case with this ratio significantly less than 
unity. Discussions of more realistic models of the noise as well 
as application of the described method to real observations will be 
done in subsequent papers.

\vspace{0.3cm}

\noi
{\bf Acknowledgments}\\
P. Naselsky and D. Novikov are grateful to the staff of TAC and the
Astronomical Observatory of the Copenhagen University for providing 
excellent working conditions during their visit to these institutions. 
This investigation was supported in part by a grant ISF MEZ 300, by a grant 
INTAS 97-1192, by the Danish Natural Science Research Council through 
grant No 9701841, by Danmarks Grundforskningsfond through its support 
for establishment of the Theoretical Astrophysics Center, by Alexander 
von Humboldt-Stiftung and by the Board of the Glasstone Benefaction.

\vspace{0.3cm}

\noi
{\large\bf References}

\vspace{0.3cm}

\noi
1. Hu., W. home page:www.sns.ias.edu/~whu/physics/physics.html

\noi
2. Tegmark., M., ApJ. Lett {\bf 480}, L87, 1997.

\noi
3. Tegmark., M. and G. Efstathiou, MNRAS {\bf 281} 1297, 1996.

\noi
4. Bond., J.R., A.N. Jaffe and L. Knox, Phys. Rev. D. {\bf 57} 2117, 1998.

\noi
5. Knox, L., astro-ph /9902046 v 2.28 Sep. 1999.

\noi
6. Taylor., A., A. Heavens, B. Ballinger and  M. Tegmark: 
      in Proceedings of the Particle Physics and Early Universe 
     Conference (PPPEUC), University of Cambridge, 7-11 April 1997.

\noi
7. Gorski, K.M., Proceedings of the XXXI-st Recontres de Marion Astrophysics 
Meeting, p.77, Editions Frontieres, (1997). astro-ph/9701191.

\noi
8. Andrews, H.C. and B.R. Hunt. Digital image restoration. New Jersey:
 Pentice-Hall, 1977.

\noi
9. Minkowski., H, Mathematische Annalen, {\bf 57}, 443, 1903.

\noi
10. Schmalzing., J. and K. Gorski, MNRAS {\bf 297}, 335, 1998.

\noi
11. Naselsky., P. and D. Novikov, ApJ., {\bf 507}, 31, 1998.

\noi
12. Novikov, D., H. Feldman, S. Shandarin, Int. Journal of Modern 
Physics D., {\bf 8} N3, 291, 1999.

\noi
13. Naselsky., P. and D. Novikov, ApJ. Lett., {\bf 444}, L1, 1996.

\noi
14. Novikov, D.I. and H.E. J\o rgensen, Int. Journal of Modern Physics D.,
 {\bf 5} N4, 319, 1996.

\noi
15. Delabrouilee, J., K. Gorski and E. Hivon, MNRAS {\bf 298}, 445, 1998.

\noi
16. Doroshkevich A.G., Astrophysics {\bf 6}, 320, 1970.

\noi
17. Press, W.H., S.A. Teukolsky, W.T. Vetteling and B.P. Flannery,
Numerical Recipes. Cambridge University Press. 1992.

\newpage

%------------------------------------Fig 1--------------------------
\begin{figure}[h]
\vspace{0.3cm}\hspace{-1cm}\epsfxsize=16cm 
\epsfbox{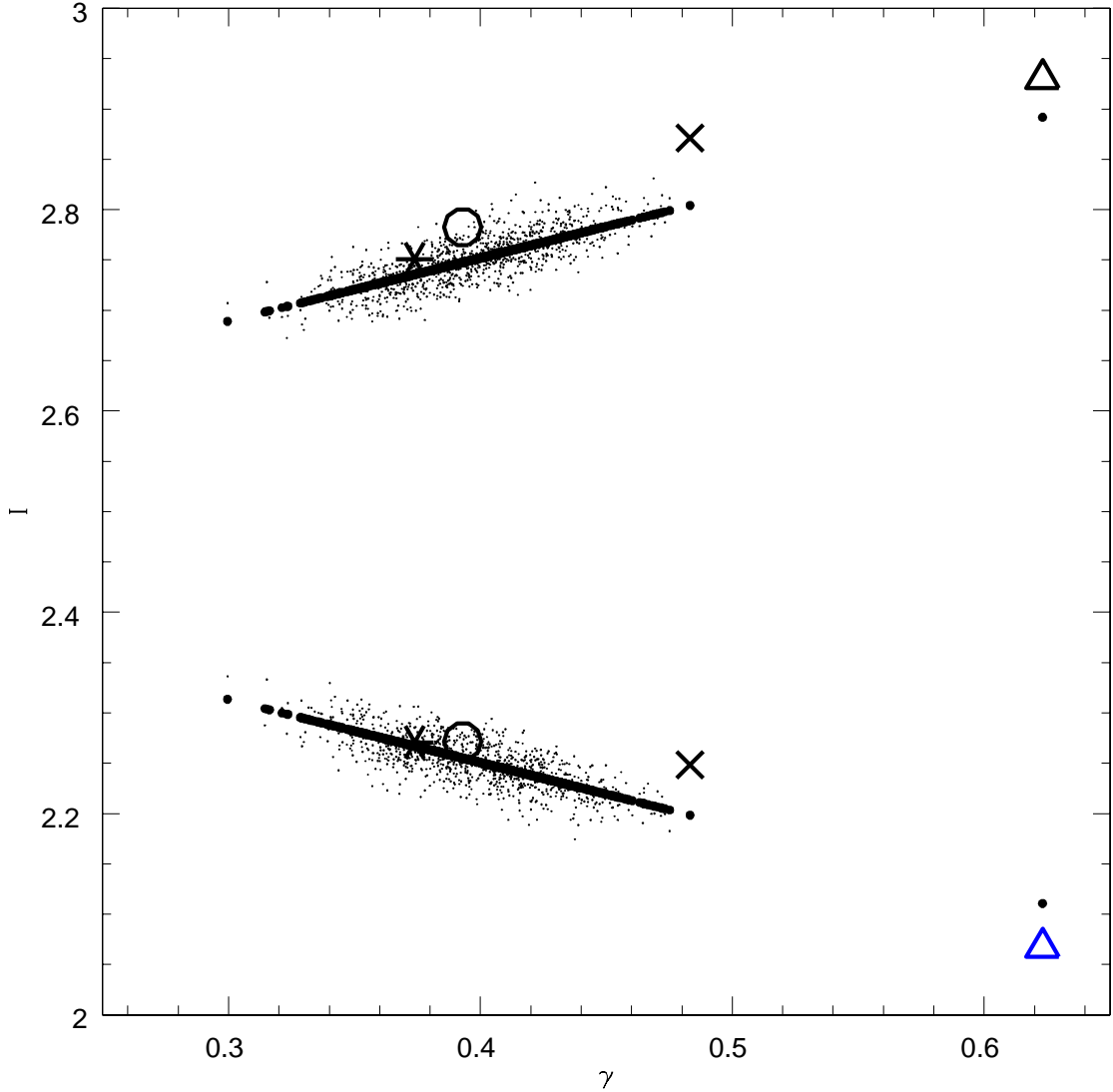}
    \caption{Dependences $I_{max}$ (upper part on vertical axis) and 
$I_{min}$ (lower part on vertical axis) of $\gamma$  (horizontal axis).
Thick dots are dependences for Gaussian analytical 
expressions (20), (24). Stars correspond to the CMB realization 
with ($\gamma$,$I_{max}$,$I_{min}$)= (0.3738, 2.751, 2.271).
Crosses correspond to the CMB+noise realization with
($\gamma$,$I_{max}$,$I_{min}$) = (0.4833, 2.871, 2.248).
Circles correspond to the reconstructed signal using the $G_p$ filter, 
parameters are ($\gamma$,$I_{max}$,$I_{min}$) = (0.3930, 2.782, 2.272). 
Triangles correspond to the reconstructed signal with the Wiener filter;
parameters are ($\gamma$,$I_{max}$,$I_{min}$) = (0.6232, 2.930, 2.067). 
}
\end{figure}
%------------------------------------Fig 1--------------------------

%------------------------------------Fig 2--------------------------
\begin{figure}[h]
\vspace{0.3cm}\hspace{-1cm}\epsfxsize=16cm 
\epsfbox{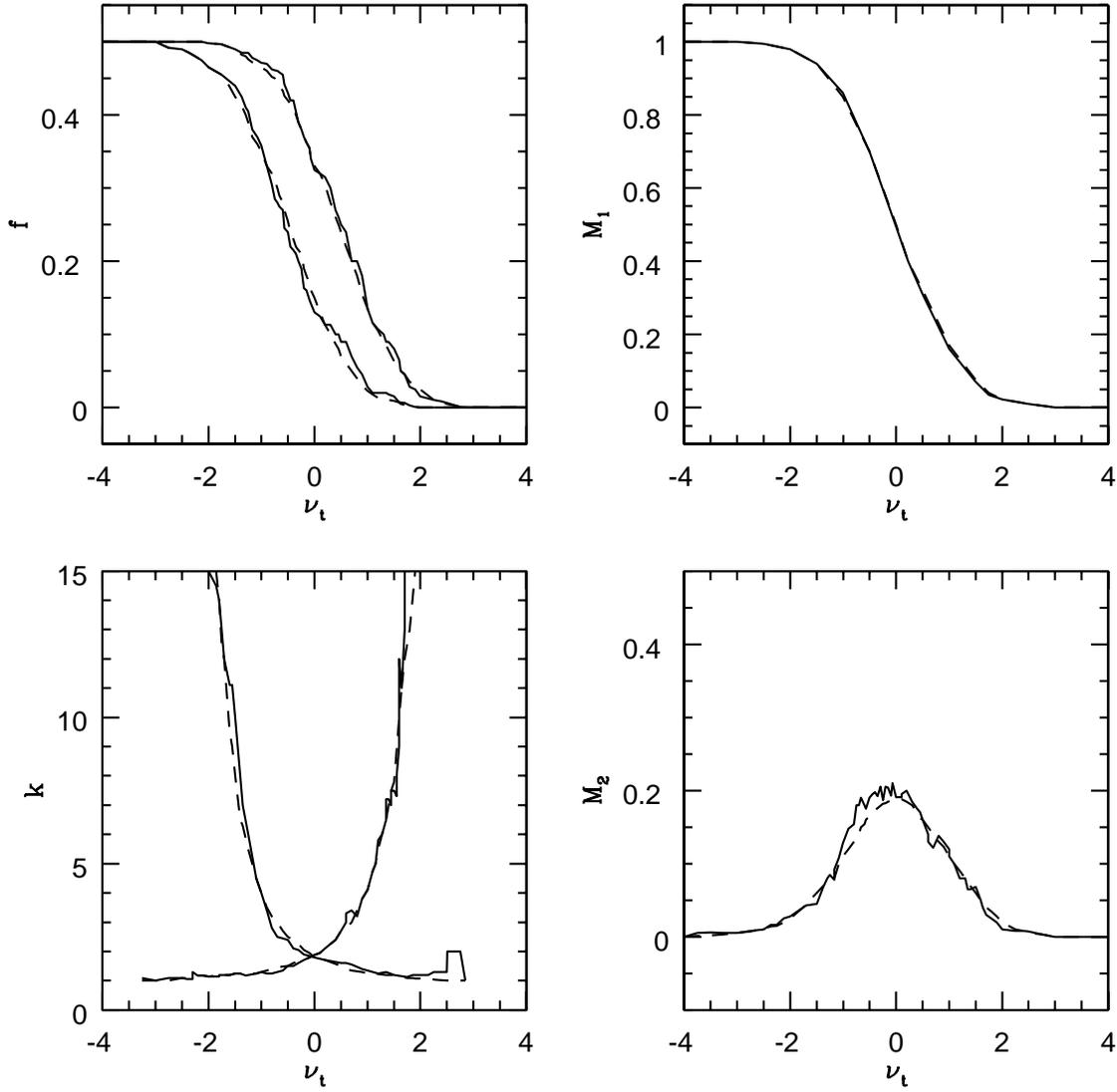}
    \caption{Dependences a) $f_{max}$, $f_{min}$, 
b) $M_1$,
c) the rate of clusterisation $\lan k\ran$ and  
d) $M_2$ as function of $\nu _t$ for a CMB realization marked on Fig. 1 by stars. 
Dashed lines correspond to analytical Gaussian expressions (see text). 
$f_{max}$ and $f_{min}$  are normalized by $\ov{N_{max}}+\ov{N_{min}}$. 
}
\end{figure}
%------------------------------------Fig 2--------------------------
%------------------------------------Fig 3--------------------------
\begin{figure}[h]
\vspace{0.3cm}\hspace{-1cm}\epsfxsize=16cm 
\epsfbox{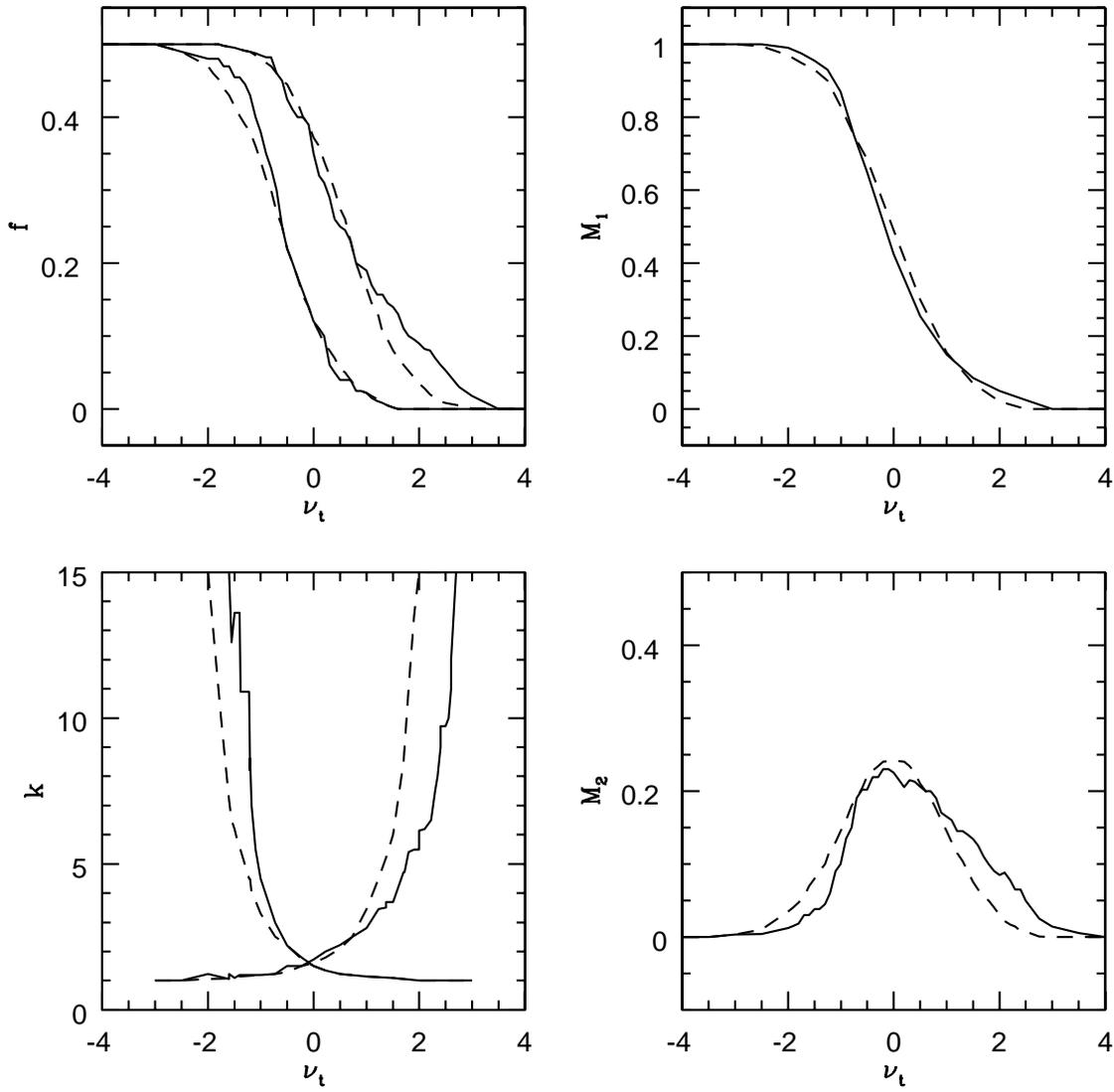}
    \caption{The same as on Fig. 2 but for the CMB+noise signal (see text).
}
\end{figure}
%------------------------------------Fig 3--------------------------
%------------------------------------Fig 4--------------------------
\begin{figure}[h]
\vspace{0.3cm}\hspace{-1cm}\epsfxsize=16cm 
\epsfbox{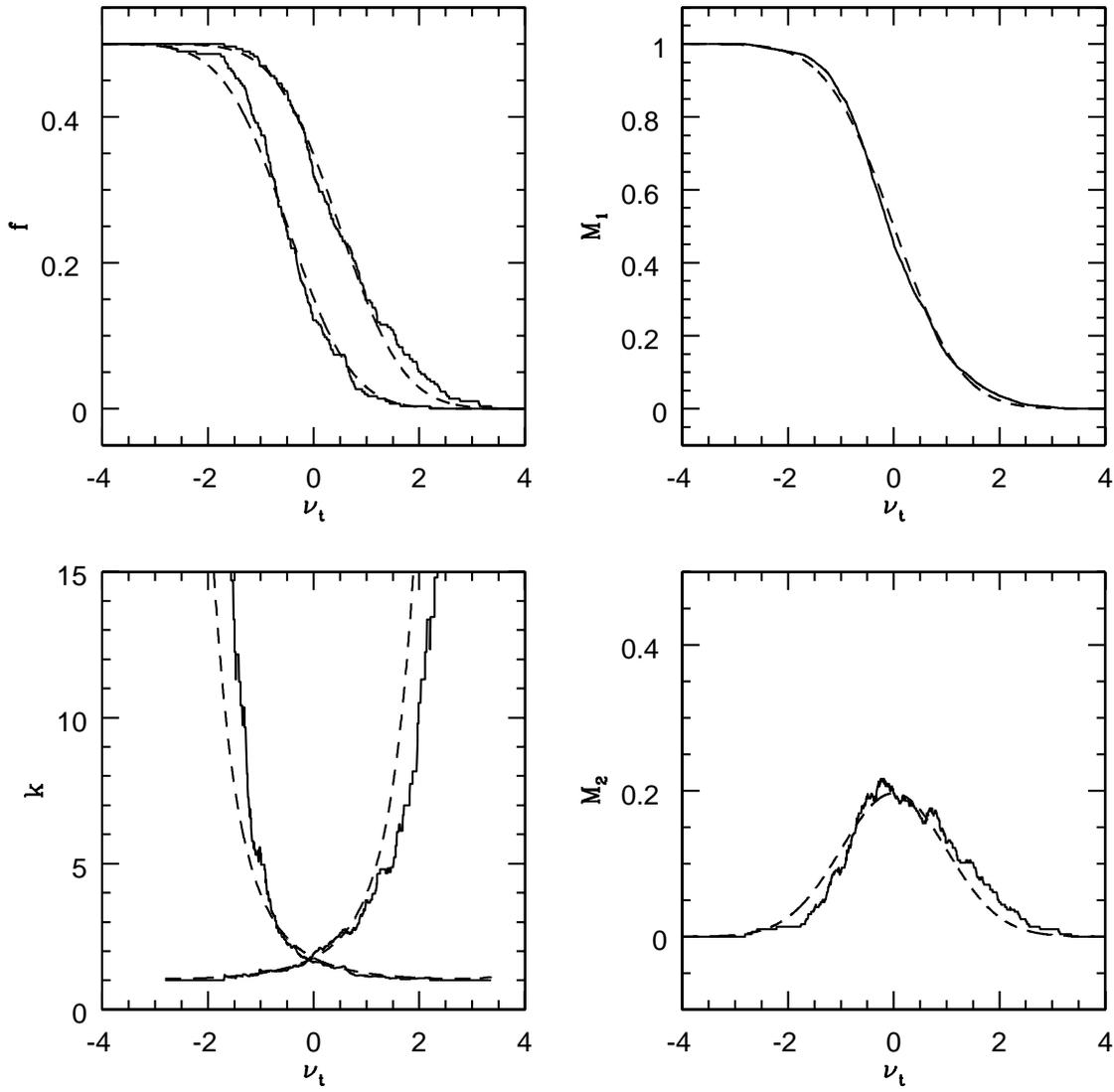}
    \caption{The same as on Fig. 2 after filtration with the $G_p$ filter (see text).
}
\end{figure}
%------------------------------------Fig 4--------------------------

\end{document}